\documentclass[apj,dvipdfm]{emulateapj}
\usepackage{graphicx,epsf,color,ulem}
\newcommand{\ltsim}{\protect\raisebox{-0.5ex}{$\:\stackrel{\textstyle <}{\sim}\:$}}
\newcommand{\gtsim}{\protect\raisebox{-0.5ex}{$\:\stackrel{\textstyle >}{\sim}\:$}}

\shorttitle{Spiral Arms in a Circumstellar Disk}
\shortauthors{Tomida et al.}
\begin{document}
\title{Grand Design Spiral Arms in a Young Forming Circumstellar Disk}

\author{Kengo Tomida\altaffilmark{1}, Masahiro N. Machida\altaffilmark{2}, Takashi Hosokawa\altaffilmark{3}, Yuya Sakurai\altaffilmark{4} and Chia Hui Lin\altaffilmark{1,5}}
\altaffiltext{1}{Department of Earth and Space Science, Osaka University, Toyonaka, Osaka, 560-0043, Japan; \mbox{tomida@vega.ess.sci.osaka-u.ac.jp}}
\altaffiltext{2}{Department of Earth and Planetary Sciences, Faculty of Sciences, Kyushu University, Nishi-ku, Fukuoka, 819-0395, Japan; \mbox{machida.masahiro.018@m.kyushu-u.ac.jp}}
\altaffiltext{3}{Department of Physics, Kyoto University, Sakyo-ku, Kyoto, 606-8502, Japan; \mbox{hosokawa@tap.scphys.kyoto-u.ac.jp}}
\altaffiltext{4}{Department of Physics, The University of Tokyo, Tokyo, 113-0033, Japan; \mbox{sakurai@utap.phys.s.u-tokyo.ac.jp}}
\altaffiltext{5}{Department of Physics, National Taiwan University, Taipei, 10617, Taiwan}

\begin{abstract}
We study formation and long-term evolution of a circumstellar disk in a collapsing molecular cloud core using a resistive magnetohydrodynamic simulation. While the formed circumstellar disk is initially small, it grows as accretion continues and its radius becomes as large as 200 AUs toward the end of the Class-I phase. A pair of grand-design spiral arms form due to gravitational instability in the disk, and they transfer angular momentum in the highly resistive disk. Although the spiral arms disappear in a few rotations as expected in a classical theory, new spiral arms form recurrently as the disk soon becomes unstable again by gas accretion. Such recurrent spiral arms persist throughout the Class-0 and I phase. We then perform synthetic observations and compare our model with a recent high-resolution observation of a young stellar object Elias 2-27, whose circumstellar disk has grand design spiral arms. We find good agreement between our theoretical model and the observation. Our model suggests that the grand design spiral arms around Elias 2-27 are consistent with material arms formed by gravitational instability. If such spiral arms commonly exist in young circumstellar disks, it implies that young circumstellar disks are considerably massive and gravitational instability is the key process of angular momentum transport.
\end{abstract}

\keywords{stars: formation --- ISM: clouds --- ISM: jets and outflows --- radiative transfer --- magnetohydrodynamics}

\section{Introduction}
Formation and evolution of a circumstellar disk is one of the most important topics in star formation, because most of the gas accretion occurs through a disk, and also because the disk provides initial and boundary conditions for planet formation. Accretion and transport of angular momentum are the key processes to understand disk formation and evolution. There are various angular momentum transport processes: gravitational torque, magnetic fields, hydrodynamic instabilities and viscosity. The relative importance of these processes depends on the physical state of the disk. In typical star forming clouds, magnetic fields are strong enough and remove angular momentum very efficiently \citep[e.g.][]{ms56,mp79,mp80,tmsk00,tmsk02}. When magnetic fields are weak or significantly dissipated by non-ideal magnetohydrodynamics (MHD) effects such as Ohmic dissipation and ambipolar diffusion, and when a disk is massive enough, spiral arms are formed by gravitational instability and the non-axisymmetric structure transports angular momentum by gravitational torque \citep[e.g.][]{bate98,tomida10b}. When these processes are inefficient, angular momentum is transported by other physical processes such as hydrodynamic instabilities and turbulence \citep[e.g.][]{turnerppvi}, which are less efficient than the other processes. These processes are not exclusive but work together. Realistic numerical simulations of star and disk formation have been actively performed \citep{dp08,li11,tomida13,tomida15,tsukamoto15,masson16,wurster16}.

Thanks to the high sensitivity and resolution of the Atacama Large Millimeter / submillimeter Array (ALMA), high resolution observations of circumstellar disks are being performed actively. Observations resolving the structures of circumstellar disks provide crucial information on the physics working in the disks. Among them, the discovery of beautiful grand design spiral arms in a circumstellar disk around Elias 2-27 \citep{e2-27} brought a striking insight on the disk evolution. Although \citet{e2-27} concluded that the spiral arms are likely to be density waves in the disk, the spiral arms look like material spiral arms formed by gravitational instability. Such spiral arms form in marginally unstable disks, where the Toomre's Q parameter \citep{toomre} is as low as a few. While the disk around Elias 2-27 seems to be considerably massive (up to $\sim 30\%$ of the central object), the estimated Q value is not very low, although this discrepancy can be reconciled by dust properties \citep[e.g.][]{tok16}. Moreover, the gravitational instability scenario is often criticized that the material arms wind up and disappear in a few dynamical timescales. It is also considered to be difficult to form such unstable disks in strongly magnetized clouds \citep[the magnetic braking catastrophe, e.g.][]{als03,hf08,li11}, at least in the early phase of star formation. If we can confirm these spiral arms are formed by gravitational instability, and if such structures are common, it suggests that angular momentum transport by gravitational torque plays a significant role in evolution of young circumstellar disks.

In this Letter, we perform a long-term resistive MHD simulation of disk formation and demonstrate that spiral arms by gravitational instability form recurrently throughout the Class-0 and I phase. We also perform synthetic observations, and show a good agreement between the observation and theoretical model. This Letter is organized as follows. We describe the MHD simulation in \S 2, and the synthetic observation in comparison with Elias 2-27 in \S 3. \S 4 is devoted for discussions and conclusions.

\section{MHD Simulation}
\subsection{Method and Model}
We perform a long-term MHD simulation using a 3-dimensional nested-grid code. We refer readers to \citet{mh13} and \citet{machida14,machida16} for the details. This code solves the MHD equations with self-gravity and Ohmic dissipation. Instead of solving expensive radiation transfer, we adopt the barotropic approximation as in \citet{mm12} to mimic the realistic thermodynamics including the initial isothermal collapse and the quasi-adiabatic first core phase \citep[e.g.][]{mi00,tomida10a,tomida13}:
\begin{equation}
T=10\, {\rm K} \left[1+\left(\frac{\rho}{\rho_{\rm crit}}\right)^{2/3}\right],
\end{equation}
where $T$ and $\rho$ are the gas temperature and density, and $\rho_{\rm crit}=3.84\times 10^{-14}\, {\rm g\, cm^{-3}}$ is the critical density where the gas becomes adiabatic.
The ionization degree $X_e$ and Ohmic resistivity $\eta$ are calculated using simple formulae \citep{nkn02,machida14} :
\begin{equation}
\eta = \frac{740}{X_e}\sqrt{\frac{T}{10\, {\rm K}}}
\end{equation}
\begin{equation}
X_e = 5.7\times 10^{-4}\left(\frac{n}{\rm cm^{-3}}\right)^{-1},
\end{equation}
where $n$ denotes the gas number density. These are fitting formulae assuming the Mathis-Rumpl-Nordsieck dust size distribution \citep{mrn} and the ionization rate is dominated by cosmic rays, $\zeta=10^{-17}\, {\rm sec^{-1}}$. This resistivity is conservative (i.e. Ohmic dissipation is weak) as shielding of cosmic rays is not included.

The initial condition is a supercritical Bonnor-Ebert-like sphere \citep{bonnor,ebert}, whose central density, temperature, radius and mass are $\rho_c = 2.2\times 10^{-18}\,{\rm g\, cm^{-3}}$, $T=10\, {\rm K}$, $R=6.1\times 10^{3}\,{\rm AU}$ and $M_0=1.25\,{\rm M_\odot}$, respectively. The sphere has a solid-body rotation around the $z$-axis with an angular velocity of $\Omega=1.5\times 10^{-13}\, {\rm sec^{-1}}$, and uniform magnetic field aligned to the rotation axis with $B_z = 36\,{\rm \mu G}$. The mass-to-flux ratio averaged over the sphere is $\mu/\mu_{\rm crit} \sim 3$. These parameters are motivated by observed star forming clouds \citep{goodman93,crutcher,hlipp6}, but NOT fine-tuned to reproduce Elias 2-27. The results do not change qualitatively even when we use different parameters.

Each nested-grid level consists of $(N_x,N_y,N_z)=(64,64,32)$ cells, with mirror symmetry imposed on the $z=0$ plane, and a finer level is created to resolve the local Jeans length with at least 8 cells \citep{trlv97}. The finest resolution achieved around the center of the computational domain is 0.75 AU, while typical resolution at the disk scale ($R\sim$ 100--200 AU) is about 3--6 AU. In order to avoid prohibitively small timesteps after formation of a protostellar core, we insert a sink particle with an accretion radius of 1 AU at the center of the domain \citep{machida14}.

\begin{figure}[t]
\scalebox{0.94}{\includegraphics{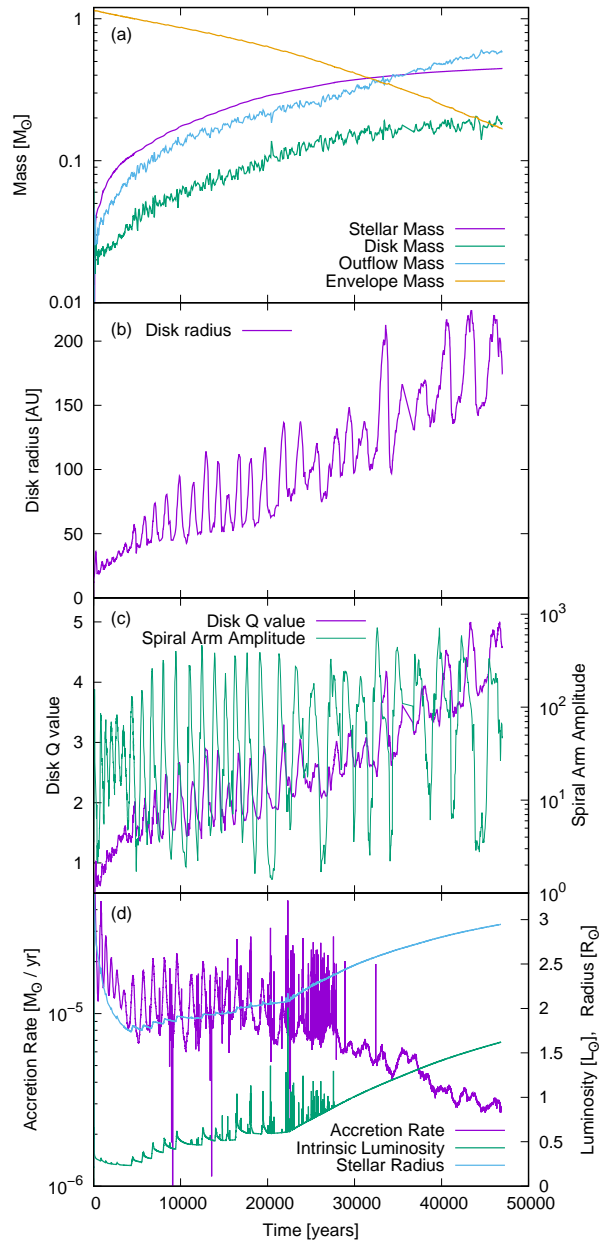}}
\caption{Time evolution of various quantities. (a): the mass of each component. (b): the disk radius. (c): Toomre's Q value averaged over the disk and the amplitude of the spiral arms. (d): The accretion rate onto the sink particle, the intrinsic luminosity and stellar radius obtained from the stellar evolution calculation. Note that the intrinsic luminosity does not include the accretion luminosity.}
\label{data}
\end{figure}

\subsection{Results}
\begin{figure*}[tp]
\scalebox{0.858}{\includegraphics{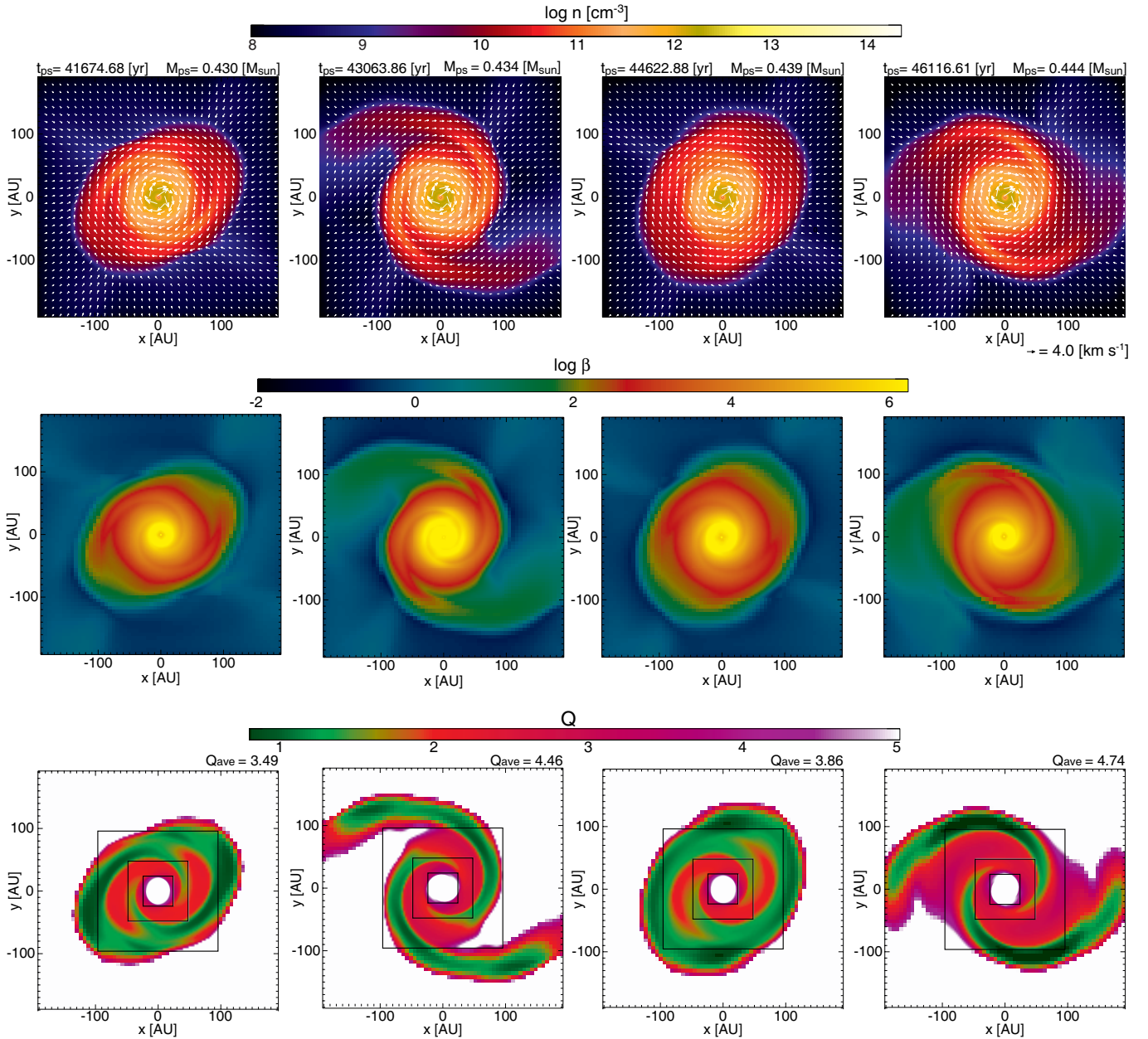}}
\caption{Typical evolution of the disk and spiral arms. Top: the density cross sections at the disk mid-plane with the velocity vectors. The supplemental movie contains both edge-on and face-on cross sections of the density at different scales. Middle: the plasma beta $\beta=p_{\rm gas}/p_{\rm mag}$ at the disk mid-plane. Bottom: the distributions of Toomre's Q parameter. We use the epoch of the rightmost panel (46,116 years after the protostar formation) for the imaging simulation.}
\label{disk}
\end{figure*}

We run the simulation as long as possible, and at the end of the simulation the protostar age reaches about 47,000 years. The masses of different components are shown in Panel (a) of Figure~\ref{data}. Here we define that the outflow ($M_{\rm outflow}$) is where the gas has a radial velocity larger than 10\% of the local isothermal sound speed ($v_r > 0.1 c_{\rm s,iso}$). The disk ($M_{\rm disk}$) is identified as the gas exceeding the critical density $\rho_{\rm d}$, which is defined as the minimum density within a region where the rotational motion dominates the radial motion ($v_\phi > 2 |v_r|$ and $v_\phi > 0.6 v_{\rm Kep}$ where $v_{\rm Kep}$ is the local Keplerian speed) and not vertically outflowing ($v_z < 0.1 c_{\rm s,iso}$). The protostar mass $M_*$ is the mass absorbed by the sink particle. The (bound) envelope mass is the total of the rest calculated within the initial cloud radius; $M_{\rm env} = M_0 - M_{*} - M_{\rm disk} - M_{\rm outflow}$. The object is almost near the end of the Class-I phase as the envelope mass gets as low as 15\% of the initial cloud mass. The disk-to-star mass ratio remains almost constant and is about 30-40\% throughout the Class-0 and I phases. Assuming that all the disk material will eventually accrete onto the star, the star formation efficiency will be $\sim 50\%$.

In the earliest phase, the disk size remains small because magnetic fields transport angular momentum efficiently. However, as the disk evolves, the magnetic angular momentum transport becomes less dominant because magnetic fields dissipate, and also because the envelope density decreases and magnetic braking becomes less efficient. Then the disk becomes gravitationally unstable, and grand-design m=2 spiral arms form spontaneously \citep[see also][]{hennebelle16}. The gravitational torque between these non-axisymmetric structures transport angular momentum efficiently and control the disk evolution. As often pointed out, these material spiral arms wind up tightly and disappear in several orbits by the shearing rotation. However, the disk becomes gravitationally unstable again and spiral arms form recurrently (see the supplement movie). While the spiral arms exist and are transferring the angular momentum, the disk radius increases and the disk becomes highly eccentric. As they transport angular momentum, the disk stabilizes and circularizes. As a result, the disk radius oscillates while it grows, as seen in Panel (b) of Figure~\ref{data}. Toward the end of the Class-I phase, the disk size becomes larger than 200 AU in the expanding phases with spiral arms.

\begin{figure}[t]
\scalebox{0.94}{\includegraphics{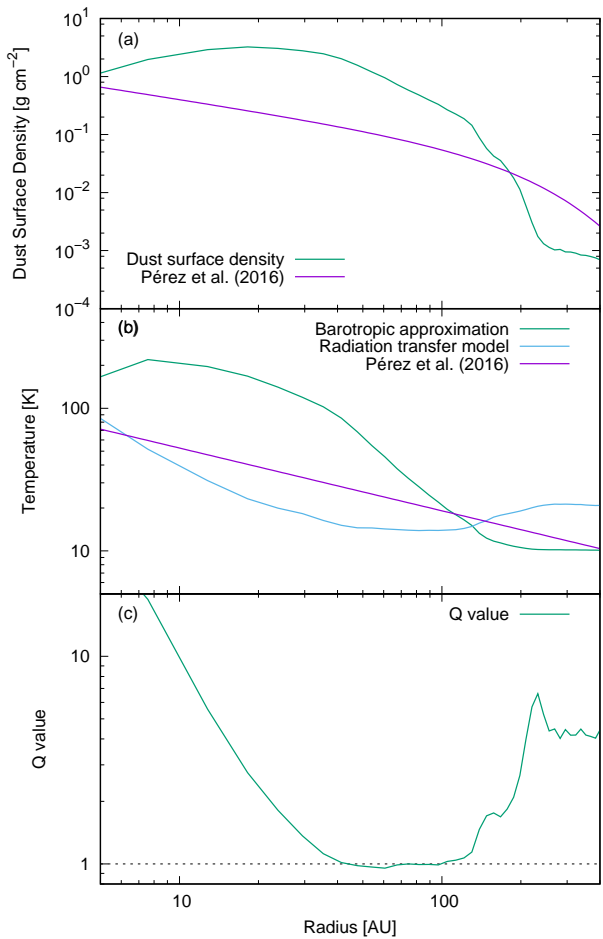}}
\caption{The azimuthally- and vertically-averaged radial distributions of various quantities. (a): the dust surface density (note that the dust-to-gas ratio is 0.014). (b): the temperature obtained from the barotropic approximation used in the MHD simulation and the recalculated temperature after radiation transfer calculation. (c): Toomre's Q value calculated using the barotropic approximation. The dashed line indicates Q=1.}
\label{rprof}
\end{figure}

In order to quantify this behavior, we plot the Q parameter \citep{toomre} averaged over the disk which indicates the disk instability, and the amplitude of the spiral arms in Panel (c) of Figure~\ref{data}. Here the Q parameter is defined as the mass-weighted average over the disk;
\begin{equation}
\langle Q \rangle = \frac{\int_{\rho>\rho_{\rm d}} \frac{c_s\kappa}{\pi G \Sigma} \Sigma dS}{\int_{\rho>\rho_{\rm d}} \Sigma dS},
\end{equation}
where $c_s$ denotes the local sound speed, $\kappa$ the epicyclic frequency, $G$ the gravitational constant and $\Sigma$ the disk surface density, respectively. We use $\kappa = \Omega_{\rm Kep}$ assuming the disk rotation is almost Keplerian. The amplitude of the spiral arms is measured by the ratio between the maximum and minimum surface densities integrated over $-50 \, {\rm AU} < z < 50 \, {\rm AU}$ within the annulus of $0.70 \, R_{\rm disk}< R< 0.75 \, R_{\rm disk}$ excluding the outflowing gas with $v_z > 0.1 c_{\rm s,iso}$. The Q value increases gradually, indicating that the disk is stabilized overall as it evolves. However, the outer disk still becomes unstable and spiral arms form repeatedly. The disk stabilizes again as the spiral arms transfer the angular momentum, which is observed as strong correlation between the spiral arm amplitude and the increase of the Q value. 

Figure~\ref{disk} shows the typical evolution of the disk, emergence and decay of the spiral arms. The disk swings between the two states in about 1,500 years; the expanding state with spiral arms and the relatively circular, flat state. The orbital timescale is $t_{\rm rot} \sim \left(\frac{4 \pi^2 R_{\rm disk}^3}{GM_*}\right)^{1/2}\sim 1,500$ years when $M_* = 0.44\, {\rm M_\odot}$ and $R_{\rm disk} = 100\, {\rm AU}$, so the transition timescale is corresponding to the disk dynamical timescale. We also show distribution of the plasma beta and the local Q value in Figure~\ref{disk}. The plasma beta is significantly higher than unity within the disk, indicating the disk is weakly magnetized and angular momentum transport by magnetic fields is not efficient any more. Spiral arms exhibit low Q ($Q \ltsim 1$) but as the disk expands, the Q value gets higher as the disk is stabilized by the angular momentum transport. Figure~\ref{rprof} shows the azimuthally averaged radial profiles of the dust surface density, temperature and Q value. Our model has a higher surface density and temperature compared to the estimates of \citet{e2-27}, but it should be noted that the assumption in \citet{e2-27} that the surface density should follow a power-law distribution with a cut-off is strong, and the disk does not necessarily have such a distribution when the disk is unstable and dynamically evolving. The Q value is as low as unity where the spiral arms are present, implying that the disk structure is regulated by the balance between the gravitational instability and the angular momentum transport by the spiral arms. As seen in Figure~\ref{data}, the recurrent spiral arms persist for a significant fraction of the time till the end of the Class-I phase. Assuming that the spiral arms are visible when the amplitude is larger than 50, the occurrence probability of the spiral arms is about 50\%. Therefore, there is sufficiently high probability that we can observe such grand-design material spiral arms if the disk is massive. Observations of Elias 2-27 suggest that its disk can be indeed massive, $M_{\rm disk}\sim 0.14\, {\rm M_\odot}$ \citep{andrews09, e2-27}.

\section{Synthetic Observation}
\begin{figure*}[tp]
\scalebox{0.2}{\includegraphics{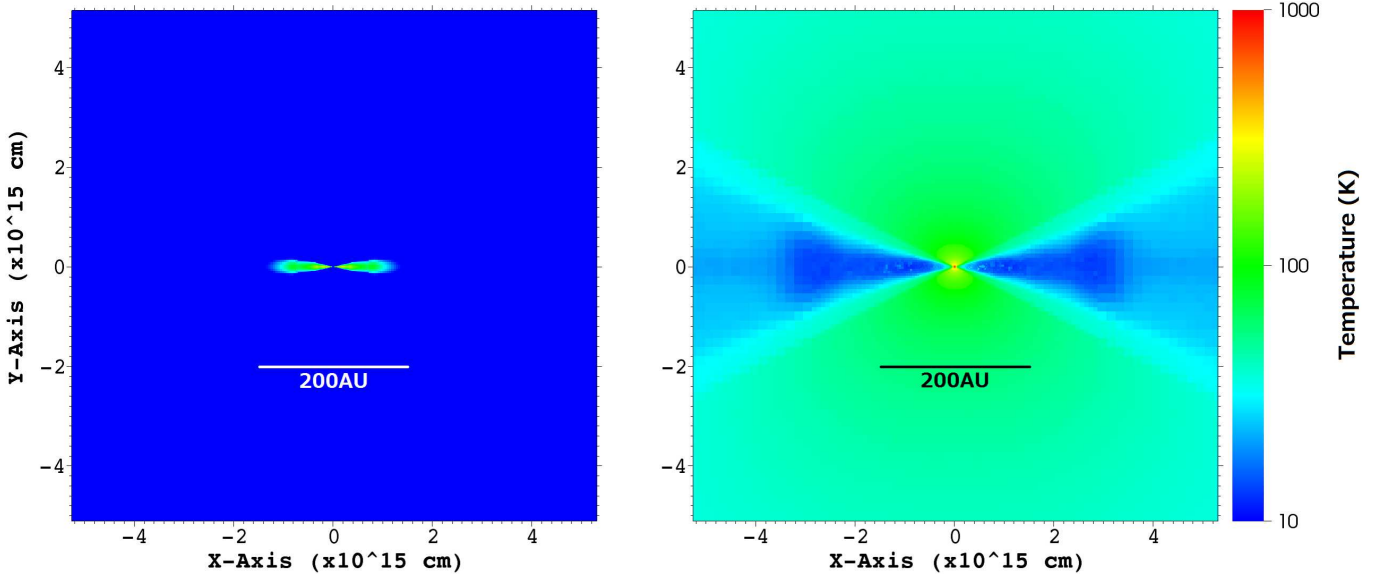}}
\caption{Vertical cross sections of the temperature. Left: the temperature calculated using the barotropic approximation. Right: the temperature after the radiation transfer calculation. The two panels share the same color bar.}
\label{temp}
\end{figure*}

In order to compare the MHD simulation with the ALMA observation \citep{e2-27}, we take a snapshot which resembles the Elias2-27 disk and perform synthetic observation to derive a dust continuum map. This procedure is done in four steps: 1) calculating the physical properties of the protostar, 2) (re)calculating the dust temperature, 3) calculating a dust continuum map, and 4) simulating an ALMA observation.

\subsection{Protostar Evolution}
We calculate evolution of the central protostar using the STELLAR stellar evolution code \citep{yb08,hosokawa13,sakurai15}, as in \citet{mh13}. Time evolution of the intrinsic luminosity and stellar radius as well as the accretion rate onto the sink particle are shown in Panel (d) of Figure~\ref{data}. At the epoch we choose, the mass, radius and intrinsic luminosity of the star are $M_*=0.444\, {\rm M_\odot}$, $R_*=2.935\, {\rm R_\odot}$ and $L_*=1.604\, {\rm L_\odot}$, respectively. The accretion rate in the MHD simulation is still high, a few $\times 10^{-6} \, {\rm M_\odot / yr}$, and the corresponding accretion luminosity should be as high as $L_{\rm acc}\sim 14\, {\rm L_\odot}$ if the gas accretion occurs steadily. However, this accretion rate is time-averaged and there can be short-time accretion variability due to small-scale unresolved structures. Both theoretically and observationally, it is known that most of young stars are fainter than theoretical estimates assuming the steady accretion rate (the so-called luminosity problem), and this is often interpreted as a result of episodic accretion. That is, while the time-averaged accretion rate is still high, the actual accretion onto the protostar happens in long quiescent phases and short burst phases triggered by some physical instabilities within the disk \citep[e.g.][]{zhu10,stamatellos12,dv12}. Considering this, we adopt $\dot{M}=8 \times 10^{-8}\, {\rm M_\odot\, yr^{-1}}$ to calculate the accretion luminosity based on the observation of Elias 2-27 \citep{najita15}, which gives $L_{\rm acc}=0.379\, {\rm L_\odot}$. Assuming a black body spectrum, the effective stellar temperature is $T_{\rm eff}=4,000\, {\rm K}$. This agrees well with the observed spectral type of Elias 2-27, which is M0 \citep{lr99,andrews09,najita15}.

\subsection{Temperature Recalculation}
While the barotropic approximation we adopt in the MHD simulation is convenient, it does not fully take account of radiation heating and cooling after formation of the protostar. To simulate the observation better, we recalculate the dust temperature distribution assuming radiative equilibrium under irradiation from the central star. For this purpose, we utilize RADMC-3D\footnote{http://www.ita.uni-heidelberg.de/\~{}dullemond/software/radmc-3d/} \citep{radmc}, which computes radiation transfer using the Monte-Carlo method. In addition to the stellar radiation, we impose an external interstellar radiation field of $T=10\,{\rm K}$.

Before the radiation transfer simulation, we modify the density distribution. First, we remove the gas and dust outside the initial cloud. Second, because the sink particle only absorbs the gas within the accretion radius whose density is higher than a critical density, $\rho_{\rm sink}=10^{-12}\, {\rm g\, cm^{-3}}$, the gas density around the sink remains unphysically high. Because the gas near the sink should already have accreted, we lower the gas density within a cylinder of $R=1.5\, {\rm AU}$ and $z=4.5\, {\rm AU}$ around the sink to $\rho=10^{-18}\, {\rm g\,cm^{-3}}$. This removal of the gas affects the temperature near the protostar, but the removed gas mass is only $5\times 10^{-5}\, {\rm M_\odot}$ and the disk temperature in the large scale ($R>10\, {\rm AU}$) is not sensitive to how we remove the gas.

For dust opacities, we adopt the monochromatic opacity tables of \citet{semenov} \footnote{http://www2.mpia-hd.mpg.de/homes/henning/Dust\_opacities /Opacities/opacities.html}, using the composite aggregate dust model of the normal abundance. The dust-to-gas fraction of this model is 0.014 in the cold ($T\ltsim 150\, {\rm K}$) region where all the dust components exist. This model consists of five tables for different temperatures regarding dust evaporation. To take the dust evaporation into account, we run RADMC-3D repeatedly. First, we run RADMC-3D using the opacity table for the lowest temperature everywhere. Based on the result, we introduce the next opacity table for the higher temperature where the temperature is higher than the evaporation temperature. We repeat this procedure until distributions of all the dust components considering their evaporation are obtained. Then we run RADMC-3D again using $2\times 10^9$ photons to make the temperature distribution as smooth as possible. We include anisotropic scattering, and do not use the modified random walk algorithm. Note that our opacity and resistivity are not fully consistent because the dust evaporation is not considered in the resistivity.

The resulting temperature distribution is shown in Figure~\ref{temp}, and the azimuthally averaged radial profile is shown in Panel (b) of Figure~\ref{rprof}. The resulting disk temperature in the outer region ($R \gtsim 100\, {\rm AU}$) is about 10--30 K, which is close to the temperature often used in observational studies. The temperature obtained from the radiation transfer calculation is lower than that of the barotropic approximation within the disk near the central star, while it is higher in the envelope. Although the temperature affects the disk stability, we expect that the overall disk evolution especially in the outer region is simulated reasonably well, because the typical temperature in the outer disk is 10--20 K, which is close to the temperature from the barotropic approximation.

The temperatures obtained from the barotropic approximation and radiation transfer calculation are considerably different, but we still can expect the gravitational instability if we properly take radiation transfer into account in our MHD simulation because the temperature from the radiation transfer calculation is lower than the barotropic approximation in the most region within the disk. Also, the Q value has a dependency on the temperature as $Q\propto T^{1/2}$ and therefore the difference in Q should be within a factor of 2. It should be noted, however, we ultimately need radiation MHD simulations to model the disk evolution accurately.

\begin{figure*}[t]
\scalebox{0.2}{\includegraphics{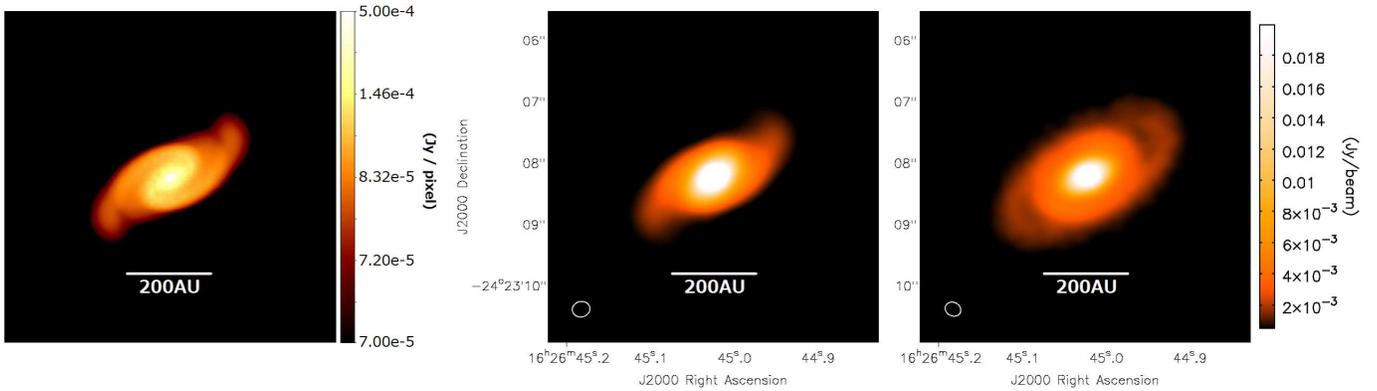}}
\caption{Left: the raw result of the RADMC-3D radiation transfer calculation. The pixel size is $(2.5 \, {\rm AU})^2$. Center: the result of the ALMA imaging simulation. The ellipse at the lower-left corner indicates the beam size of $0.29\, {\rm arcsec} \times 0.26\, {\rm arcsec}$. Right: the actual ALMA observation of Elias 2-27 \citep[][https://safe.nrao.edu/evla/disks/elias2-27/]{e2-27}. The two right panels share the same color bar.}
\label{sim}
\end{figure*}
\subsection{Dust Continuum Map}
For comparison with the ALMA observation of Elias 2-27, we calculate the intensity distribution of dust continuum at 1.3 mm again using RADMC-3D. The opacity table we use has $\kappa_{1.3{\rm mm}} \sim 0.95 \, {\rm cm^2 \, g^{-1}}$ per unit dust mass in most region within the disk. Note that this value is lower than the opacity used in \citet{e2-27} by a factor of $\sim 2.5$. To match the observation, we set the inclination angle to be 55.8 degrees. In order to avoid unphysical noise in the ALMA simulation due to the edges of the image, the image size must be larger than the primary beam size of ALMA, which is about 27 arcsec at 1.3 mm. Assuming the object is located at a distance of 139 pc, we calculate the image covering $(10,000\, {\rm AU})^2$ with $4,000^2$ pixels. The result is shown in the left panel of Figure~\ref{sim}. 

The column density along the line of the sight from the edge to the center of the cloud, excluding the gas within the sink radius, is $N_{\rm H_2} \sim 6.34 \times 10^{21} \, {\rm cm^{-2}}$, indicating that the envelope is almost exhausted. This value is lower than one inferred from the observed visual extinction $A_V\sim 14.8$ \citep{andrews09,najita15}. However, the column density estimate in our simulation is only a lower limit because we exclude the structure near the sink and outside the initial cloud. 

\subsection{ALMA Simulation}

Then we perform simulated observation using the {\it simobserve} and {\it simanalyze} tasks of the Common Astronomy Software Applications package \citep[CASA,][]{casa}. We chose observation parameters as close to \citet{e2-27} as possible; the bandwidth is 6.8 GHz, the integration time is 12.5 minutes, and the source position is Right Ascension (J2000) = 16h26m45.024s, Declination (J2000) = -24d23m08.250s. We use ``alma.cycle4.5.cfg" array configuration file and the Briggs waiting with the robustness of 0.5. This produces a synthetic beam of $0.29\, {\rm arcsec} \times 0.26\, {\rm arcsec}$, which is very close to the original observation. The result of the synthetic observation is shown in Figure~\ref{sim}, as well as the original observation of Elias 2-27 \citep{e2-27}.

The disk structure agrees well with the ALMA observation of Elias 2-27. The central bright disk and the grand-design spiral arms are clearly visible. The disk size, brightness, and thickness of the spiral arms are largely consistent with the observation. Therefore, we conclude that the observed spiral arms can be well explained by material spiral arms formed by gravitational instability.

\section{Discussions and Conclusions}
We perform 3D resistive MHD simulations of star and disk formation for a long term until almost the end of the main accretion phase. The disk becomes gravitationally unstable and grand-design spiral arms form recurrently on the dynamical timescale. We compare our model with the observation of Elias 2-27 through the synthetic observation, and find a good agreement. While \citet{e2-27} favored the density wave scenario pointing out that material spiral arms should disappear in a few orbital periods, we conclude that material spiral arms that form recurrently by gravitational instability can well explain the observed grand-design spiral arms in Elias 2-27.

The current MHD simulation does not consider ambipolar diffusion, the Hall effect and radiation transfer. Ambipolar diffusion and the Hall effect will suppress angular momentum transport more effectively and help formation of the spiral arms. While we perform post-processing radiation transfer to recalculate the temperature, this is also not necessarily accurate because it ignores heating processes such as compression, shock and Joule heating. We need long-term radiation MHD simulations in order to study the disk evolution accurately, which will be a future work.

Our results imply that Elias 2-27 is very young, and is possibly in the earliest Class-II or very close to the end of the Class-I phase. Although the stellar evolution calculation has a significant uncertainty for such a young object, \citet{isella09} \citep[see also][]{gm95,aw07} estimated that the age of Elias 2-27 is very young, about 0.1 Myrs. At least this is one of the youngest Class-II objects, and our model is qualitatively consistent with the observations. Such a short duration of the Class-0/I phase is due to the high accretion rate; while the self-similar solution of \citet{shu77} predicts $\dot{M} \sim 10^{-6} \, {\rm M_\odot / yr}$, the actual accretion is more like the Larson-Penston solution and the accretion rate is much higher in the early phase \citep{lrs69,pen69,vaytet16}. One remaining question is how long the recurrent spiral arms can last. As the envelope disperses and the accretion rates decreases, the circumstellar disk will be stabilized. This trend is already seen in our simulation. Therefore, we speculate that the disk will be stabilized shortly after the accretion from the envelope stops.

While our model reproduces the disk of Elias 2-27 well, some features do not match perfectly. For example, gap-like structures exist in Elias 2-27 but we do not find any significant gap in our synthetic observation. The low-density region between the spiral arms may look like the gap if we use higher resolution and contrast. However, we should discuss the gap-like structures carefully because the observed contrast is very low. Also, while a bipolar molecular outflow is launched in our model (see the supplemental movie), no significant outflow is detected in the small scale. In our simulation, the molecular outflow is launched near the disk in the early phase, but the launching region moves to the higher envelope above the disk ($z\sim 250\, {\rm AU}$ at the end of the simulation) in the late phase because the disk is not strongly magnetized any more and the magnetic field configuration changes. Also, the outflow is getting weaker as the accretion rate decreases. Therefore, if Elias 2-27 is slightly more evolved than our model, it is likely that there is no outflow in the small scale, as observed in \citet{e2-27}. In the scale of the molecular cloud core or larger, on the other hand, \citet{gurney08} reported detection of low velocity (1-2 km/s) molecular outflows toward Elias 2-27. This velocity is in a good agreement with our MHD simulation. We are planning to perform synthetic observations of molecular lines and compare with observations in a future work.

It has been pointed out that theoretical models and numerical simulations tend to predict circumstellar disks heavier than the observed Class-0/I objects, while massive disks with $M_{\rm disk}>0.1\, {\rm M_\odot}$ do exist \citep{jor09,tobin16}. This ``disk-mass problem" can be reconciled if massive disks with spiral arms exist commonly. Such massive disks are favorable to explain the observed high binary rate \citep[e.g.][]{dk13} based on the disk fragmentation scenario. Moreover, it is important to consider such massive disks in the context of planet formation, because circumstellar disks are the initial and boundary conditions of planet formation. Systematic survey of young circumstellar disks are of crucial importance, along with improvement of theoretical models.

\acknowledgments
We thank Takayuki Muto, Akimasa Kataoka and Laura P{\'e}rez for useful information and discussion. We also thank Cornelis P. Dullemond for publicly distributing RADMC-3D. We use VisIt \citep{visit} to produce Figure~\ref{temp}. VisIt is supported by the Department of Energy with funding from the Advanced Simulation and Computing Program and the Scientific Discovery through Advanced Computing Program. This paper makes use of the following ALMA data: ADS/JAO.ALMA\#2013.1.00498.S. ALMA is a partnership of ESO (representing its member states), NSF (USA) and NINS (Japan), together with NRC (Canada), NSC and ASIAA (Taiwan), and KASI (Republic of Korea), in cooperation with the Republic of Chile. The Joint ALMA Observatory is operated by ESO, AUI/NRAO and NAOJ. This work is partly supported by the Ministry of Education, Culture, Sports, Science and Technology (MEXT), Grants-in-Aid for Scientific Research 16H05998(KT), 25400232(MNM), 16H05996(HT) and Grant-in-Aid for the Japan Society for the Promotion of Science Fellows 15H08816(YS). YS is also supported by Advanced Leading Graduate Course for Photon Science. This research uses computational resources of the High Performance Computing Infrastructure (HPCI) system provided by Cyber Sciencecenter (Tohoku University), Cybermedia Center (Osaka University) and Earth Simulator (JAMSTEC) through the HPCI System Research Project (Project ID:hp160079).


\begin{thebibliography}{}
\expandafter\ifx\csname natexlab\endcsname\relax\def\natexlab#1{#1}\fi

\bibitem[{{Allen} {et~al.}(2003){Allen}, {Li}, \& {Shu}}]{als03}
{Allen}, A., {Li}, Z.-Y., \& {Shu}, F.~H. 2003, \apj, 599, 363

\bibitem[{{Andrews} \& {Williams}(2007)}]{aw07}
{Andrews}, S.~M., \& {Williams}, J.~P. 2007, \apj, 671, 1800

\bibitem[{{Andrews} {et~al.}(2009){Andrews}, {Wilner}, {Hughes}, {Qi}, \&
  {Dullemond}}]{andrews09}
{Andrews}, S.~M., {Wilner}, D.~J., {Hughes}, A.~M., {Qi}, C., \& {Dullemond},
  C.~P. 2009, \apj, 700, 1502

\bibitem[{{Bate}(1998)}]{bate98}
{Bate}, M.~R. 1998, \apjl, 508, L95

\bibitem[{{Bonnor}(1956)}]{bonnor}
{Bonnor}, W.~B. 1956, \mnras, 116, 351

\bibitem[{Childs {et~al.}(2012)Childs, Brugger, Whitlock, Meredith, Ahern,
  Pugmire, Biagas, Miller, Harrison, Weber, Krishnan, Fogal, Sanderson, Garth,
  Bethel, Camp, R\"{u}bel, Durant, Favre, \& Navr\'{a}til}]{visit}
Childs, H., Brugger, E., Whitlock, B., {et~al.} 2012, in {High Performance
  Visualization--Enabling Extreme-Scale Scientific Insight}, 357--372

\bibitem[{{Crutcher}(2012)}]{crutcher}
{Crutcher}, R.~M. 2012, \araa, 50, 29

\bibitem[{{Duch{\^e}ne} \& {Kraus}(2013)}]{dk13}
{Duch{\^e}ne}, G., \& {Kraus}, A. 2013, \araa, 51, 269

\bibitem[{{Duffin} \& {Pudritz}(2008)}]{dp08}
{Duffin}, D.~F., \& {Pudritz}, R.~E. 2008, \mnras, 391, 1659

\bibitem[{{Dullemond}(2012)}]{radmc}
{Dullemond}, C.~P. 2012, {RADMC-3D: A multi-purpose radiative transfer tool},
  Astrophysics Source Code Library, ascl:1202.015

\bibitem[{{Dunham} \& {Vorobyov}(2012)}]{dv12}
{Dunham}, M.~M., \& {Vorobyov}, E.~I. 2012, \apj, 747, 52

\bibitem[{{Ebert}(1955)}]{ebert}
{Ebert}, R. 1955, Zeitschrift fur Astrophysik, 36, 222

\bibitem[{{Goodman} {et~al.}(1993){Goodman}, {Benson}, {Fuller}, \&
  {Myers}}]{goodman93}
{Goodman}, A.~A., {Benson}, P.~J., {Fuller}, G.~A., \& {Myers}, P.~C. 1993,
  \apj, 406, 528

\bibitem[{{Greene} \& {Meyer}(1995)}]{gm95}
{Greene}, T.~P., \& {Meyer}, M.~R. 1995, \apj, 450, 233

\bibitem[{{Gurney} {et~al.}(2008){Gurney}, {Plume}, \& {Johnstone}}]{gurney08}
{Gurney}, M., {Plume}, R., \& {Johnstone}, D. 2008, \pasp, 120, 1193

\bibitem[{{Hennebelle} {et~al.}(2016){Hennebelle}, {Commer{\c c}on},
  {Chabrier}, \& {Marchand}}]{hennebelle16}
{Hennebelle}, P., {Commer{\c c}on}, B., {Chabrier}, G., \& {Marchand}, P. 2016,
  \apjl, 830, L8

\bibitem[{{Hennebelle} \& {Fromang}(2008)}]{hf08}
{Hennebelle}, P., \& {Fromang}, S. 2008, \aap, 477, 9

\bibitem[{{Hosokawa} {et~al.}(2013){Hosokawa}, {Yorke}, {Inayoshi}, {Omukai},
  \& {Yoshida}}]{hosokawa13}
{Hosokawa}, T., {Yorke}, H.~W., {Inayoshi}, K., {Omukai}, K., \& {Yoshida}, N.
  2013, \apj, 778, 178

\bibitem[{{Isella} {et~al.}(2009){Isella}, {Carpenter}, \&
  {Sargent}}]{isella09}
{Isella}, A., {Carpenter}, J.~M., \& {Sargent}, A.~I. 2009, \apj, 701, 260

\bibitem[{{J{\o}rgensen} {et~al.}(2009){J{\o}rgensen}, {van Dishoeck},
  {Visser}, {Bourke}, {Wilner}, {Lommen}, {Hogerheijde}, \& {Myers}}]{jor09}
{J{\o}rgensen}, J.~K., {van Dishoeck}, E.~F., {Visser}, R., {et~al.} 2009,
  \aap, 507, 861

\bibitem[{{Larson}(1969)}]{lrs69}
{Larson}, R.~B. 1969, \mnras, 145, 271

\bibitem[{{Li} {et~al.}(2014){Li}, {Goodman}, {Sridharan}, {Houde}, {Li},
  {Novak}, \& {Tang}}]{hlipp6}
{Li}, H.-b., {Goodman}, A., {Sridharan}, T.~K., {et~al.} 2014, ArXiv e-prints,
  arXiv:1404.2024

\bibitem[{{Li} {et~al.}(2011){Li}, {Krasnopolsky}, \& {Shang}}]{li11}
{Li}, Z.-Y., {Krasnopolsky}, R., \& {Shang}, H. 2011, \apj, 738, 180

\bibitem[{{Luhman} \& {Rieke}(1999)}]{lr99}
{Luhman}, K.~L., \& {Rieke}, G.~H. 1999, \apj, 525, 440

\bibitem[{{Machida} \& {Hosokawa}(2013)}]{mh13}
{Machida}, M.~N., \& {Hosokawa}, T. 2013, \mnras, 431, 1719

\bibitem[{{Machida} {et~al.}(2014){Machida}, {Inutsuka}, \&
  {Matsumoto}}]{machida14}
{Machida}, M.~N., {Inutsuka}, S.-i., \& {Matsumoto}, T. 2014, \mnras, 438, 2278

\bibitem[{{Machida} \& {Matsumoto}(2012)}]{mm12}
{Machida}, M.~N., \& {Matsumoto}, T. 2012, \mnras, 421, 588

\bibitem[{{Machida} {et~al.}(2016){Machida}, {Matsumoto}, \&
  {Inutsuka}}]{machida16}
{Machida}, M.~N., {Matsumoto}, T., \& {Inutsuka}, S.-i. 2016, \mnras, 463, 4246

\bibitem[{{Masson} {et~al.}(2016){Masson}, {Chabrier}, {Hennebelle}, {Vaytet},
  \& {Commer{\c c}on}}]{masson16}
{Masson}, J., {Chabrier}, G., {Hennebelle}, P., {Vaytet}, N., \& {Commer{\c
  c}on}, B. 2016, \aap, 587, A32

\bibitem[{{Masunaga} \& {Inutsuka}(2000)}]{mi00}
{Masunaga}, H., \& {Inutsuka}, S.-i. 2000, \apj, 531, 350

\bibitem[{{Mathis} {et~al.}(1977){Mathis}, {Rumpl}, \& {Nordsieck}}]{mrn}
{Mathis}, J.~S., {Rumpl}, W., \& {Nordsieck}, K.~H. 1977, \apj, 217, 425

\bibitem[{{McMullin} {et~al.}(2007){McMullin}, {Waters}, {Schiebel}, {Young},
  \& {Golap}}]{casa}
{McMullin}, J.~P., {Waters}, B., {Schiebel}, D., {Young}, W., \& {Golap}, K.
  2007, in Astronomical Society of the Pacific Conference Series, Vol. 376,
  Astronomical Data Analysis Software and Systems XVI, ed. R.~A. {Shaw},
  F.~{Hill}, \& D.~J. {Bell}, 127

\bibitem[{{Mestel} \& {Spitzer}(1956)}]{ms56}
{Mestel}, L., \& {Spitzer}, Jr., L. 1956, \mnras, 116, 503

\bibitem[{{Mouschovias} \& {Paleologou}(1979)}]{mp79}
{Mouschovias}, T.~C., \& {Paleologou}, E.~V. 1979, \apj, 230, 204

\bibitem[{{Mouschovias} \& {Paleologou}(1980)}]{mp80}
---. 1980, \apj, 237, 877

\bibitem[{{Najita} {et~al.}(2015){Najita}, {Andrews}, \&
  {Muzerolle}}]{najita15}
{Najita}, J.~R., {Andrews}, S.~M., \& {Muzerolle}, J. 2015, \mnras, 450, 3559

\bibitem[{{Nakano} {et~al.}(2002){Nakano}, {Nishi}, \& {Umebayashi}}]{nkn02}
{Nakano}, T., {Nishi}, R., \& {Umebayashi}, T. 2002, \apj, 573, 199

\bibitem[{{Penston}(1969)}]{pen69}
{Penston}, M.~V. 1969, \mnras, 144, 425

\bibitem[{{P{\'e}rez} {et~al.}(2016){P{\'e}rez}, {Carpenter}, {Andrews},
  {Ricci}, {Isella}, {Linz}, {Sargent}, {Wilner}, {Henning}, {Deller},
  {Chandler}, {Dullemond}, {Lazio}, {Menten}, {Corder}, {Storm}, {Testi},
  {Tazzari}, {Kwon}, {Calvet}, {Greaves}, {Harris}, \& {Mundy}}]{e2-27}
{P{\'e}rez}, L.~M., {Carpenter}, J.~M., {Andrews}, S.~M., {et~al.} 2016,
  Science, 353, 1519

\bibitem[{{Sakurai} {et~al.}(2015){Sakurai}, {Hosokawa}, {Yoshida}, \&
  {Yorke}}]{sakurai15}
{Sakurai}, Y., {Hosokawa}, T., {Yoshida}, N., \& {Yorke}, H.~W. 2015, \mnras,
  452, 755

\bibitem[{{Semenov} {et~al.}(2003){Semenov}, {Henning}, {Helling}, {Ilgner}, \&
  {Sedlmayr}}]{semenov}
{Semenov}, D., {Henning}, T., {Helling}, C., {Ilgner}, M., \& {Sedlmayr}, E.
  2003, \aap, 410, 611

\bibitem[{{Shu}(1977)}]{shu77}
{Shu}, F.~H. 1977, \apj, 214, 488

\bibitem[{{Stamatellos} {et~al.}(2012){Stamatellos}, {Whitworth}, \&
  {Hubber}}]{stamatellos12}
{Stamatellos}, D., {Whitworth}, A.~P., \& {Hubber}, D.~A. 2012, \mnras, 427,
  1182

\bibitem[{{Tobin} {et~al.}(2016){Tobin}, {Kratter}, {Persson}, {Looney},
  {Dunham}, {Segura-Cox}, {Li}, {Chandler}, {Sadavoy}, {Harris}, {Melis}, \&
  {P{\'e}rez}}]{tobin16}
{Tobin}, J.~J., {Kratter}, K.~M., {Persson}, M.~V., {et~al.} 2016, \nat, 538,
  483

\bibitem[{{Tomida} {et~al.}(2010{\natexlab{a}}){Tomida}, {Machida}, {Saigo},
  {Tomisaka}, \& {Matsumoto}}]{tomida10b}
{Tomida}, K., {Machida}, M.~N., {Saigo}, K., {Tomisaka}, K., \& {Matsumoto}, T.
  2010{\natexlab{a}}, \apjl, 725, L239

\bibitem[{{Tomida} {et~al.}(2015){Tomida}, {Okuzumi}, \& {Machida}}]{tomida15}
{Tomida}, K., {Okuzumi}, S., \& {Machida}, M.~N. 2015, \apj, 801, 117

\bibitem[{{Tomida} {et~al.}(2013){Tomida}, {Tomisaka}, {Matsumoto}, {Hori},
  {Okuzumi}, {Machida}, \& {Saigo}}]{tomida13}
{Tomida}, K., {Tomisaka}, K., {Matsumoto}, T., {et~al.} 2013, \apj, 763, 6

\bibitem[{{Tomida} {et~al.}(2010{\natexlab{b}}){Tomida}, {Tomisaka},
  {Matsumoto}, {Ohsuga}, {Machida}, \& {Saigo}}]{tomida10a}
---. 2010{\natexlab{b}}, \apjl, 714, L58

\bibitem[{{Tomisaka}(2000)}]{tmsk00}
{Tomisaka}, K. 2000, \apjl, 528, L41

\bibitem[{{Tomisaka}(2002)}]{tmsk02}
---. 2002, \apj, 575, 306

\bibitem[{{Toomre}(1964)}]{toomre}
{Toomre}, A. 1964, \apj, 139, 1217

\bibitem[{{Truelove} {et~al.}(1997){Truelove}, {Klein}, {McKee}, {Holliman},
  {Howell}, \& {Greenough}}]{trlv97}
{Truelove}, J.~K., {Klein}, R.~I., {McKee}, C.~F., {et~al.} 1997, \apjl, 489,
  L179

\bibitem[{{Tsukamoto} {et~al.}(2015){Tsukamoto}, {Iwasaki}, {Okuzumi},
  {Machida}, \& {Inutsuka}}]{tsukamoto15}
{Tsukamoto}, Y., {Iwasaki}, K., {Okuzumi}, S., {Machida}, M.~N., \& {Inutsuka},
  S. 2015, \apjl, 810, L26

\bibitem[{{Tsukamoto} {et~al.}(2016){Tsukamoto}, {Okuzumi}, \&
  {Kataoka}}]{tok16}
{Tsukamoto}, Y., {Okuzumi}, S., \& {Kataoka}, A. 2016, ArXiv e-prints,
  arXiv:1608.03015

\bibitem[{{Turner} {et~al.}(2014){Turner}, {Fromang}, {Gammie}, {Klahr},
  {Lesur}, {Wardle}, \& {Bai}}]{turnerppvi}
{Turner}, N.~J., {Fromang}, S., {Gammie}, C., {et~al.} 2014, Protostars and
  Planets VI, 411

\bibitem[{{Vaytet} \& {Haugb{\o}lle}(2016)}]{vaytet16}
{Vaytet}, N., \& {Haugb{\o}lle}, T. 2016, ArXiv e-prints, arXiv:1610.03324

\bibitem[{{Wurster} {et~al.}(2016){Wurster}, {Price}, \& {Bate}}]{wurster16}
{Wurster}, J., {Price}, D.~J., \& {Bate}, M.~R. 2016, \mnras, 457, 1037

\bibitem[{{Yorke} \& {Bodenheimer}(2008)}]{yb08}
{Yorke}, H.~W., \& {Bodenheimer}, P. 2008, in Astronomical Society of the
  Pacific Conference Series, Vol. 387, Massive Star Formation: Observations
  Confront Theory, ed. H.~{Beuther}, H.~{Linz}, \& T.~{Henning}, 189

\bibitem[{{Zhu} {et~al.}(2010){Zhu}, {Hartmann}, {Gammie}, {Book}, {Simon}, \&
  {Engelhard}}]{zhu10}
{Zhu}, Z., {Hartmann}, L., {Gammie}, C.~F., {et~al.} 2010, \apj, 713, 1134

\end{thebibliography}
\end{document}